\documentclass[prc,reprint,longbibliography,amsmath,showpacs]{revtex4-1}

\usepackage{graphicx}
\usepackage[utf8]{inputenc} 
\usepackage{color}
\usepackage{upgreek}
\usepackage{amssymb}
\usepackage{fdsymbol}

\newcommand{\un}[1]{{\ensuremath{\,\text{#1}}}}

\newcommand{\subscript}[1]{\ensuremath{_{\textrm{#1}}}}   % subscript for bibfile

\begin{document}

%some abbrevs.
\newcommand{\e}{\mathrm{e}} %eulersche Zahl
\newcommand{\im}{\mathrm{i}} %imaginäre Zahl
\newcommand{\Vb}{V_{\mathrm{sd}}}
\newcommand{\Vg}{V_{\mathrm{g}}}
\newcommand{\EB}{E_{\mathrm{B}}}

%Chemical formulas
\newcommand{\He}{$^{3}\mathrm{He}$/$^{4}\mathrm{He}$ }
\newcommand{\SiO}{$\text{SiO}_{2}$ }
\newcommand{\CH}{$\text{CH}_{4}$}
\newcommand{\Hyd}{$\text{H}_{2}$}
\newcommand{\ReMo}[2]{\ensuremath{\text{Mo}_{#2}\text{Re}_{#1}}}
\newcommand{\MoRe}[2]{\ensuremath{\text{Mo}_{#1}\text{Re}_{#2}}}

%dimension L to put in $$
\newcommand{\nm}{\, \text{nm} } %Nanometer 
\newcommand{\mikrom}{\, \upmu\text{m} } %Mukrometer 
\newcommand{\mm}{\, \text{mm} } %Millimeter
\newcommand{\cm}{\, \text{cm} } %Centimeter

%pressure unit to put in $$
\newcommand{\mbar}{\un{mbar}}

%Superconductivity put in $$
\newcommand{\Tc}{\ensuremath{T_{\text{c}}}} %crit. temp
\newcommand{\dR}{\ensuremath{\text{d}R}} %diff res.
\newcommand{\Ic}{\ensuremath{I_{\text{c}}}}	%crit. curr
\newcommand{\Bc}{\ensuremath{B_{\text{c}}}}	%crit. b field
\newcommand{\Ir}{\ensuremath{I_{\text{r}}}} %retrapping curr
\newcommand{\jc}{\ensuremath{j_{\text{c}}}} %critical current density

%resonator stuff
\newcommand{\fr}{\ensuremath{f_{\text{r}}}} % resonance frequency
\newcommand{\Qi}{\ensuremath{Q_{\text{i}}}} % internal q
\newcommand{\Qe}{\ensuremath{Q_{\text{e}}}} % external q
\newcommand{\Qc}{\ensuremath{Q_{\text{c}}}} % capacitival q
\newcommand{\Ql}{\ensuremath{Q_{\text{l}}}} % load q

\newcommand{\kB}{\ensuremath{k_{\text{B}}}}

\title{Co-sputtered MoRe thin films for carbon nanotube growth-compatible 
superconducting coplanar resonators}

\author{K. J. G. Götz}
\email{karl.goetz@ur.de}
\author{S. Blien}
\author{P. L. Stiller}
\author{O. Vavra}
\author{T. Mayer}
\author{T. Huber}
\author{T. N. G. Meier}
\author{M. Kronseder}
\author{Ch. Strunk}
\author{A. K. Hüttel}
\email{andreas.huettel@ur.de}
\affiliation{Institute for Experimental and Applied Physics, University of
Regensburg, Universitätsstr.\ 31, D-93053 Regensburg, Germany}

\begin{abstract}
Molybdenum rhenium alloy thin films can exhibit superconductivity up to 
critical temperatures of $\Tc=15\un{K}$. At the same time, the films are highly 
stable in the high-temperature methane / hydrogen atmosphere typically required 
to grow single wall carbon nanotubes. We characterize molybdenum rhenium alloy 
films deposited via simultaneous sputtering from two sources, with respect to 
their composition as function of sputter parameters and their electronic dc as 
well as GHz properties at low temperature. Specific emphasis is placed on 
the effect of the carbon nanotube growth conditions on the film. 
Superconducting coplanar waveguide resonators are defined lithographically; we 
demonstrate that the resonators remain functional when undergoing nanotube 
growth conditions, and characterize their properties as function of 
temperature. This paves the way for ultra-clean nanotube devices grown in situ 
onto superconducting coplanar waveguide circuit elements.
\end{abstract}


\maketitle

\section{Introduction}

The discovery of single-walled carbon nanotubes \cite{Iijima1993, Bethune1993} 
opened the door for a wide range of both fundamental research and technical 
applications in such electronic and nano-electromechanical systems. These 
macromolecules exhibit particular properties as, e.g., high mechanical, 
chemical and thermal stability as well as excellent electronic and thermal 
conductance. They can carry high currents, display outstanding room 
temperature transistor characteristics, and may compete in the future with or 
even replace conventional silicon-based logic devices \cite{Avouris2002, 
Avouris2007, Shulaker2013}. From a fundamental point of view carbon nanotubes 
provide a highly promising material for quantum information and computation. 
The macromolecules act as quasi one-dimensional conductors because of their 
large aspect ratio; the electrostatic definition of localized potential wells 
can trap electrons in a nuclear spin free environment with well-defined quantum 
levels \cite{GroveRasmussen2007, SteeleDQD2009, Benyamini2014, brokensu4, 
splitkondo}.

One way to ensure that the carbon nanotubes remain clean and defect-free is to 
grow them directly on pre-fabricated contact electrodes, such that subsequently 
no further lithography and / or wet chemistry takes place. This technique has 
led to a multitude of striking measurement results, from single quantum dot 
spectroscopy \cite{nmat-cao:745, brokensu4} to the characterization of high-$Q$ 
vibrational modes \cite{highq, mosermillion}.
\begin{figure}[t]
\includegraphics[width=\columnwidth]{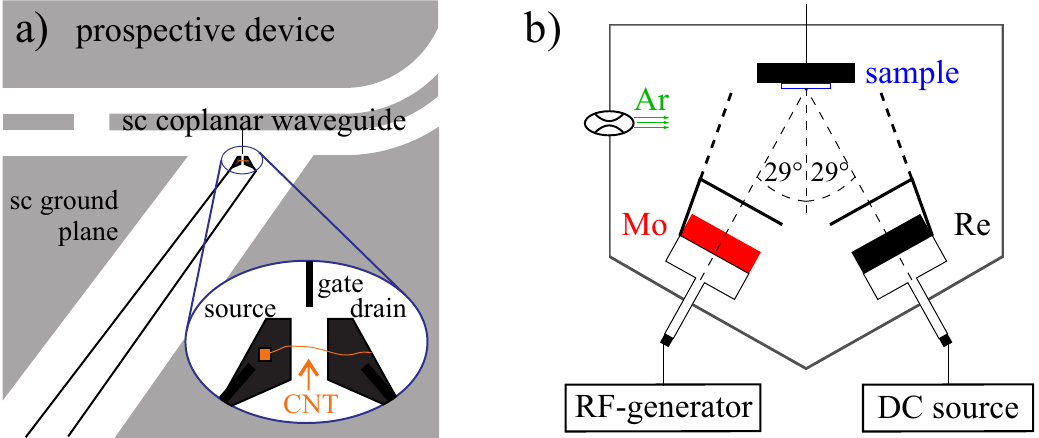} 
\caption{(Color online) (a) Schematic sketch of a carbon nanotube quantum dot 
device capacitively coupled to a coplanar $\lambda/2$ resonator. For similar 
devices using different material systems, see Refs.~\onlinecite{Delbecq2011, 
Viennot2014}. (b) Sketch of the Orion dual-source sputtering setup used during 
fabrication of our metal thin films. The molybdenum target is driven by a 
radiofrequency source with impedance matching, the rhenium target using a dc 
source. A mass flow controller sets the argon flow into chamber.} 
\label{fig:cosp}
\end{figure}
Performing high-frequency measurements on single wall carbon nanotube 
structures is a logical extension. Many techniques applied for such 
measurements on mesoscopic structures require the use of on-chip radiofrequency 
electronics, in particular superconducting coplanar waveguides and microwave 
resonators formed by these \cite{Viennot2014, Ranjan2015}. An example for such 
a possible combined device structure is sketched in Fig. \ref{fig:cosp}(a) -- a 
carbon nanotube is capacitively coupled to the coplanar $\lambda/2$ resonator. 
In combination with the overgrowth process, this poses extensive difficulties, 
since the carbon nanotube growth process --- typically $10$-$15\un{min}$ in a 
methane--hydrogen athmosphere \cite{growth} --- is highly detrimental to 
superconducting thin films.  

Several solutions for this problem have been proposed, including transfer of the 
nanotube from a growth chip to a second structure \cite{StampingWu, 
StampingDelft, Benyamini2014, Ranjan2015, Viennot2014} or capping the 
superconductor with a protection layer \cite{Delbecq2011}. If the metal film is 
to survive the CVD conditions, a high melting point, as provided by e.g. rhenium 
and molybdenum, becomes an important requirement.

Alloys of rhenium and molybdenum have been shown to exhibit superconducting 
transition temperatures up to $15 \un{K}$ \cite{EnTcMoRe, A15MoRe, 
EbeamSCMoRe}. Rhenium and molybdenum rhenium alloy thin films remain stable 
under carbon nanotube CVD growth conditions and subsequently still exhibit 
superconducting behavior \cite{SchneiderSCCNT}. In addition, after in-situ 
growth they provide transparent electronic contacts to carbon nanotubes 
\cite{SchneiderSCCNT, brokensu4, pssb-stiller:2518}.

In the following, we characterize molybdenum rhenium alloy films deposited via 
simultaneous sputtering from two sources, i.e., cosputtering, with respect to 
their composition as function of sputter parameters and their dc as well as GHz 
properties at low temperature. Specific emphasis is placed on the effect of the 
carbon nanotube growth conditions on the film. Coplanar waveguide $\lambda/4$ 
resonators are defined lithographically and characterized at dilution 
refrigerator temperatures; even after undergoing the growth conditions internal 
quality factors of up to $\Qi \simeq 5000$ can be found. The temperature 
dependence of the resonance frequency and the internal quality factor is 
evaluated and found consistent with theoretical models.

\section{Thin film deposition}
Figure \ref{fig:cosp}(b) sketches the setup used for the film fabrication. The 
UHV-chamber with typical pressures in the range of $10^{-8}$ to $10^{-7} 
\un{mbar}$ contains two sputter targets. The sample holder is placed 
approximately $13.5 \un{cm}$ above and in the middle of both targets which are 
mounted at a distance of circa $15\un{cm}$ to each other. Argon gas is injected 
via a mass flow controller close to the molybdenum target, and a plasma is 
ignited using a radiofrequency drive at a chamber pressure of 
$10^{-1}\un{mbar}$.   

Subsequently the chamber pressure is reduced to $7\cdot 10^{-3} \un{mbar}$ and 
the argon plasma at the rhenium electrode is ignited using a dc power supply; at 
the same time the rf-output is adjusted such that the plasma close to the 
molybdenum electrode and target remains stable. As soon as both targets are 
sputtered, the shutters are opened and the deposition of the MoRe alloy starts. 
By keeping the chamber pressure constant over the whole process time, the 
deposition rate of both materials is kept approximately constant. In the 
following the alloy composition of the films is varied only by tuning one 
parameter, namely the output power $P_{\text{Mo}}$ of the rf-generator at the 
molybdenum target.  

\section{Resulting alloy}
In order to obtain the alloy composition, X-ray photoelectron spectroscopy 
(XPS) is performed on the cosputtered films \cite{XPSMoRe, XPSTcMoRe}. Within a 
predefined area of four to ten square millimeters the samples are irradiated by 
a monochromatic X-ray source, and the resulting emitted photoelectrons are 
collected and spectroscopically analyzed with respect to energy and intensity. 
The chemical sensitivity is given by the element-specific distribution of 
binding energies $E_{\text{B}}$ of the electronic core levels.
\begin{figure}[t]
\includegraphics[width=\columnwidth]{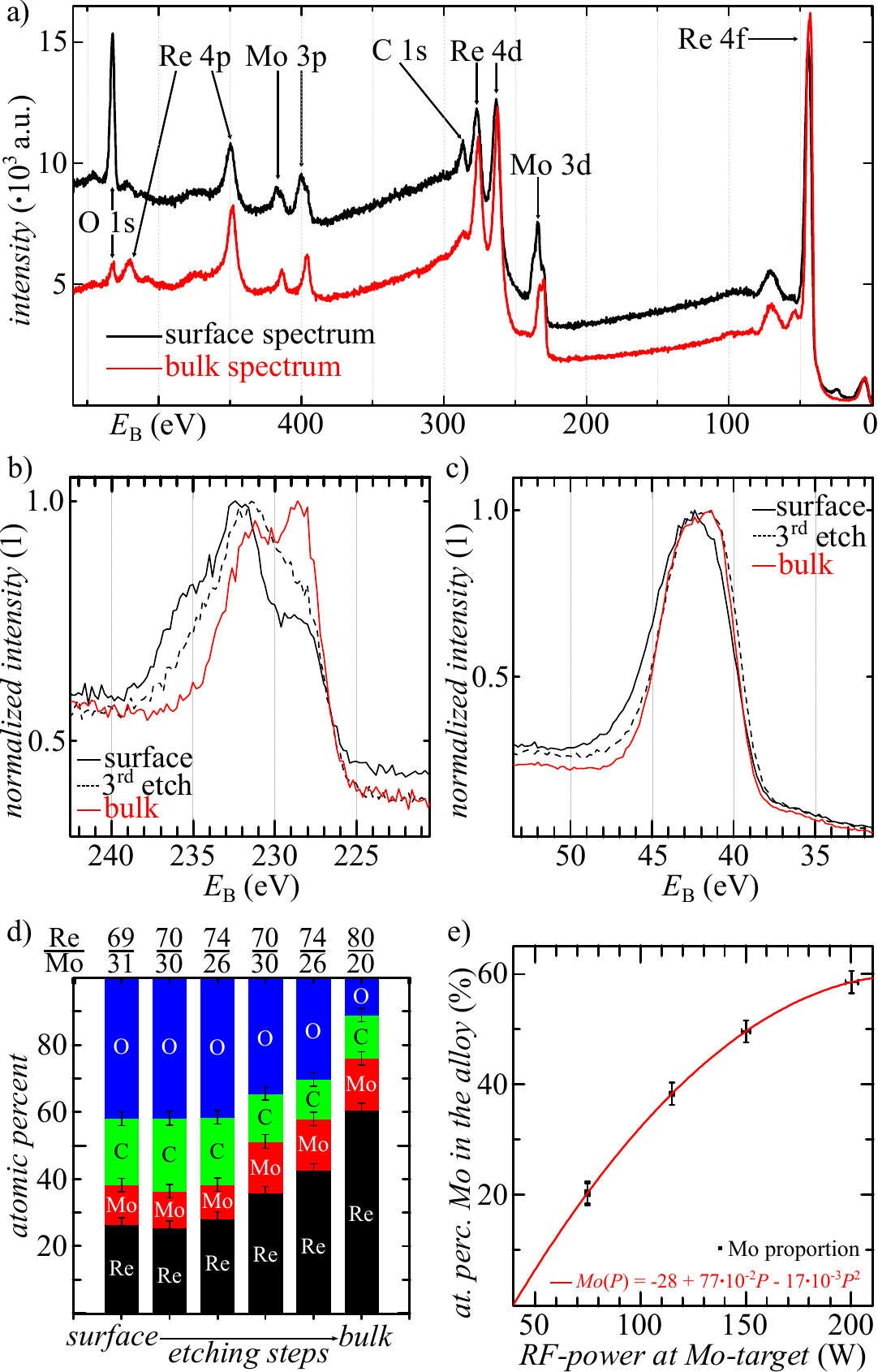} 
\caption{(Color online) (a) Example X-ray photoelectron spectroscopy (XPS) 
spectrum of a deposited, unstructured film on a $\text{p}^{++}$-Si/\SiO 
substrate. A power setting of $P_{\text{Mo}}=75\un{W}$ at the molybdenum source 
was used. The black line represents the chemical composition of the as-grown 
sample surface, the red line was recorded after removing approximately 
$5\un{nm}$ of the film by in-situ argon-ion sputtering beam sputtering within 
the XPS chamber. Normalized detail spectra close to the $\text{Mo} \, 
3\text{d}$-peaks (b) and the $\text{Re} \, 4\text{f}$-peaks (c) (see text). The 
energetic step size is set to $0.2 \un{eV}$. (d) Depth profile of the atomic 
composition, calculated from XPS spectra as in (a), from surface to $\sim 
5\un{nm}$ depth (subsequently named bulk). The resulting MoRe-alloy composition 
after each step is stated above. (e) Relative molybdenum atomic percentage in 
the bulk film for different power settings $P_{\text{Mo}}$ at the molybdenum 
target (see text).} 
\label{fig:xpseich}
\end{figure}

Characteristic XPS data of a cosputtered film are plotted in 
Fig.~\ref{fig:xpseich}(a). The signal peaks corresponding to atomic and molecular 
core levels have been identified following Ref.~\onlinecite{XPShandbook}. 
The black curve in Fig.~\ref{fig:xpseich}(a) reveals the chemical composition of 
the topmost surface layers. This is due to the small mean free path of the 
photoexcited electrons within the film material of few nanometers. In addition 
to characteristic peaks originating from molybdenum and rhenium core levels, 
also significant oxygen and carbon peaks are observed. This is likely due to 
the fact that all samples have been exposed to air during transfer from the 
sputtering device into the UHV chamber of the XPS setup. 

Adsorbates and the film itself can be etched by {\it in situ} Ar-sputtering for 
obtaining spectra from lower layers. An etching step lasts several minutes at a 
chamber pressure of less than $3 \cdot 10^{-8} \mbar$ in which the base pressure 
is of $5 \cdot 10^{-10} \mbar$. Etching steps are repeated until the oxygen 
1s-peak only negligibly contributes to the whole spectrum (at most about 10\%). 
The red (gray) line in Fig.~\ref{fig:xpseich}(a) displays the corresponding 
spectrum. Note that also the carbon peak is now strongly suppressed. Subsequent 
tests using a profilometer show that approximately $5\un{nm}$ of the films were 
removed. All spectra obtained after such a corresponding etching step are 
henceforth denoted as ``bulk'' spectra. 

Figure \ref{fig:xpseich}(b) and \ref{fig:xpseich}(c) display details of the 
spectrum close to the $\text{Mo} \, 3\text{d}$- and $\text{Re} \, 4\text{f}$-
peaks, referenced to the $\text{C}\, 1\text{s}$-peak.\cite{XPSTcMoRe}  
The additional structure in the $\text{Mo} \, 3\text{d}$ surface spectrum 
originates from $\text{MoO}_{3}$ and $\text{MoO}_{2}$ forming at the alloy 
surface \cite{MoOxiChem, XPSMoRe}. It decreases subsequently until 
disappearing in the bulk where only the two $\text{Mo} \, 3\text{d}_{5/2}$- and 
$\text{Mo} \, 3\text{d}_{3/2}$-peaks are part of the spectrum. In contrast, at 
the $\text{Re} \, 4\text{f}$-peak neither changes in the line shape nor peak 
shifts are observed in the surface compared to the bulk spectrum, indicating 
the absence of rhenium oxides at the surface and in the bulk layers. All films 
examined in this work exhibit the same behavior for the rhenium peaks 
\cite{XPSTcMoRe, XPSMoRe}.

All rhenium peaks of the film are shifted by $1.3 \un{eV}$ to higher 
$E_{\text{B}}$-values, compared to the literature values of pure rhenium.
Similar values have been identified in Ref.~\onlinecite{XPSTcMoRe} as chemical 
shift due to the Mo-Re compound formation.

Using the method of area sensitivity factors and evaluating the $\text{Mo} 
\, 3\text{d}$-, $\text{Re} \, 4\text{f}$-, $\text{C} \, 1\text{s}$-, and 
$\text{O} \, 1\text{s}$-peaks, the atomic concentrations of the sample have 
been estimated \cite{XPShandbook}.  Fig.~\ref{fig:xpseich}(d) displays the 
atomic concentrations as a function of depth. While the exposure to air leads 
to significant carbon and oxygen percentages at the surfaces, these both 
decrease strongly; for the ``bulk'' spectrum about 10\% of carbon remains. 

Subsequently the molybdenum-rhenium alloy ratio is obtained by normalizing to 
the sum of both sputtered metals. As can be clearly seen in 
Fig.~\ref{fig:xpseich}(e), where the resulting alloy ratio in the bulk film 
is plotted as a function of rf power $P_{\text{Mo}}$ applied to the molybdenum 
sputter source, the resulting molybdenum contribution in the alloy can be 
controlled over a wide range. The solid line in Fig.~\ref{fig:xpseich}(e) is a 
quadratic fit to the data points.

\section{Influence of the nanotube growth environment}

\begin{figure}[t]
\includegraphics[width=\columnwidth]{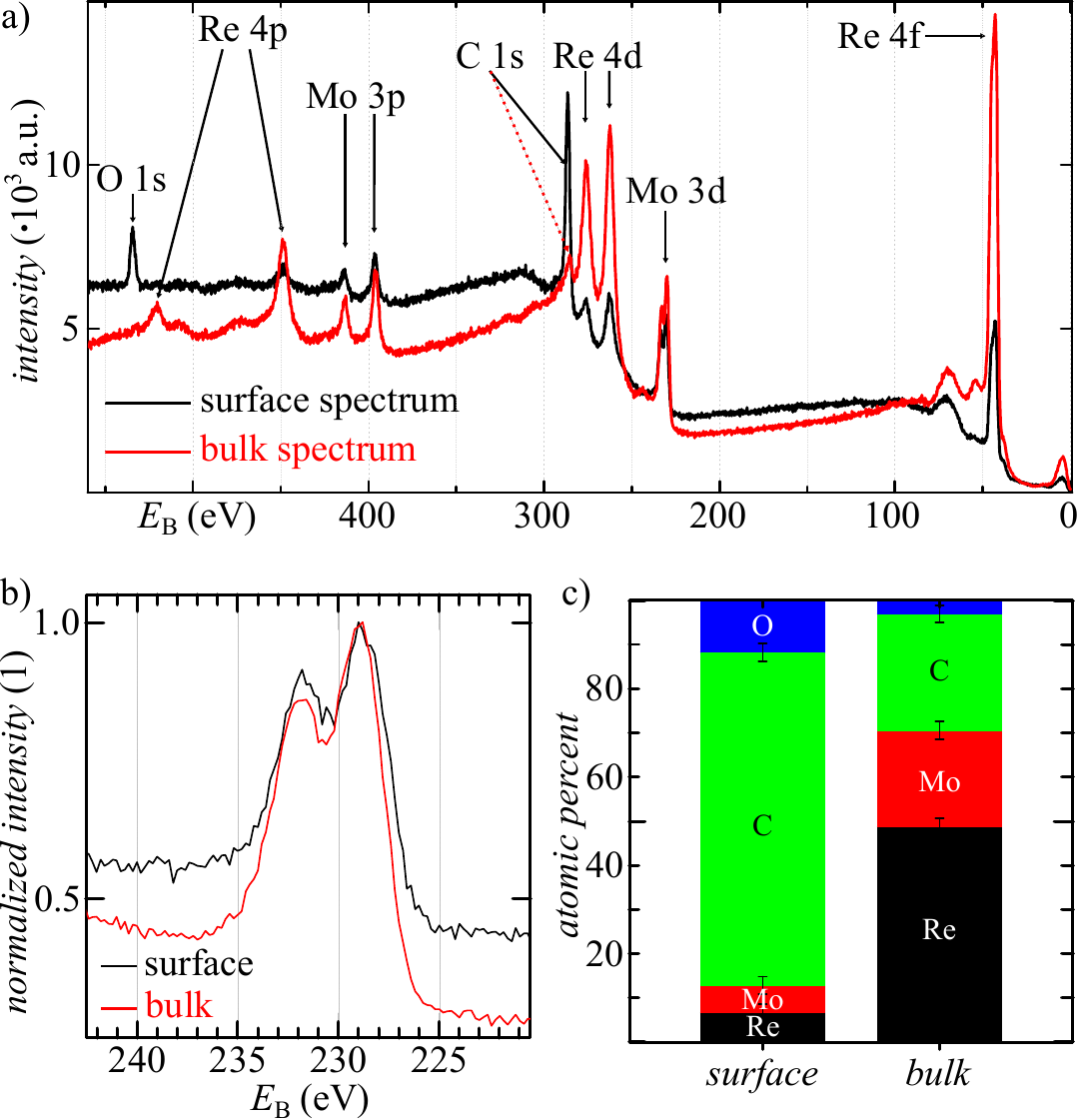} 
\caption{(Color online) (a) XPS spectrum of a molybdenum rhenium film sputtered 
at $P_{\text{Mo}}=75\un{W}$, after subsequent $30\un{min}$ in the carbon 
nanotube CVD growth environment. The energy step size is set to $0.2\un{eV}$. 
(b) Detailed zoom of the $\text{Mo} \, 3\text{d}$ peak. (c) Composition of the 
film at the surface and at $\sim 5\un{nm}$ depth (atomic percentage); note the 
large carbon contribution.}
\label{fig:3}
\end{figure}
Simulating the carbon nanotube chemical vapor deposition (CVD) growth process 
typically used to locally grow few clean single-wall carbon nanotubes 
\cite{growth}, the chip including the MoRe thin film is heated up in an argon 
and hydrogen gas flow, then exposed to a methane-hydrogen atmosphere at about 
$900\,^\circ\text{C}$ for several minutes and subsequently cooled down under 
argon and hydrogen flow.

The XPS of a molybdenum rhenium alloy sputtered with $P_{\text{Mo}} = 
75 \un{W}$ after exposure to the CVD environment is shown in 
Fig.~\ref{fig:3}(a), and a normalized detail plot of the $\text{Mo} \, 
3\text{d}$ peak in Fig.~\ref{fig:3}(b). A strong carbon peak is visible even in 
the bulk. Furthermore in both surface and bulk molybdenum spectra oxide peaks 
are not observed, as visible in Fig.~\ref{fig:xpseich}(b). 

Having in mind that molybdenum oxides exhibit drastically lower melting points 
of $795\, ^\circ\text{C}$ for $\text{MoO}_{3}$, and $1100 \,^\circ\text{C}$ for 
$\text{MoO}_{2}$ \cite{Hollemann} than pure molybdenum with $2610\, 
^\circ\text{C}$ \cite{Stoecker}, the data indicates that reduction of the oxide 
to atomic molybdenum takes place during CVD. 

An additionally possible process is the formation of molybdenum carbides. The 
expected XPS peak of $\text{Mo}_{2}\text{C}$ ($\EB=227.75 \un{eV}$) is very 
close to that of $\text{Mo} \, 3\text{d}_{5/2}$  ($\EB=228 \un{eV}$), which 
makes detection challenging with our experimental resolution. Interestingly, 
molybdenum carbides display superconductivity with critical temperatures 
$6\un{K}\lesssim \Tc \lesssim 9\un{K}$ \cite{Mo2CTc}. However, typically the 
growth of molybdenum carbide out of pure molybdenum, for example in a 
nitrogen-xylene atmosphere, takes place at temperatures higher than $900 
\,^\circ \text{C}$, see \cite{Mo2CTc}. 

The area sensitivity factor analyzing method results in a bulk composition 
of 26\% carbon, 22\% molybdenum, and 49\% rhenium plus neglectable oxygen 
residues. Since the etching time for the bulk spectra has been kept constant, 
it is obvious that the high carbon contribution does not have its origin in 
atmospheric adsorbates but in diffusion of carbon into the alloy during CVD. 
The relative atomic ratio of the sputtered metals is \ReMo{69}{31}, indicating 
structural changes during CVD. 

Spectroscopy on a second sample using twice the gas flow of methane results in a 
composition of 12\% molybdenum, 23\% rhenium, and even 58\% carbon in the bulk. 
From this we conclude that the penetration of carbon into the alloy also 
increases. However, the resulting sputtered metal ratio of \ReMo{66}{34} still 
remains close to the previous sample.

\section{dc characterization}

\begin{figure}[t]
\includegraphics[width=\columnwidth]{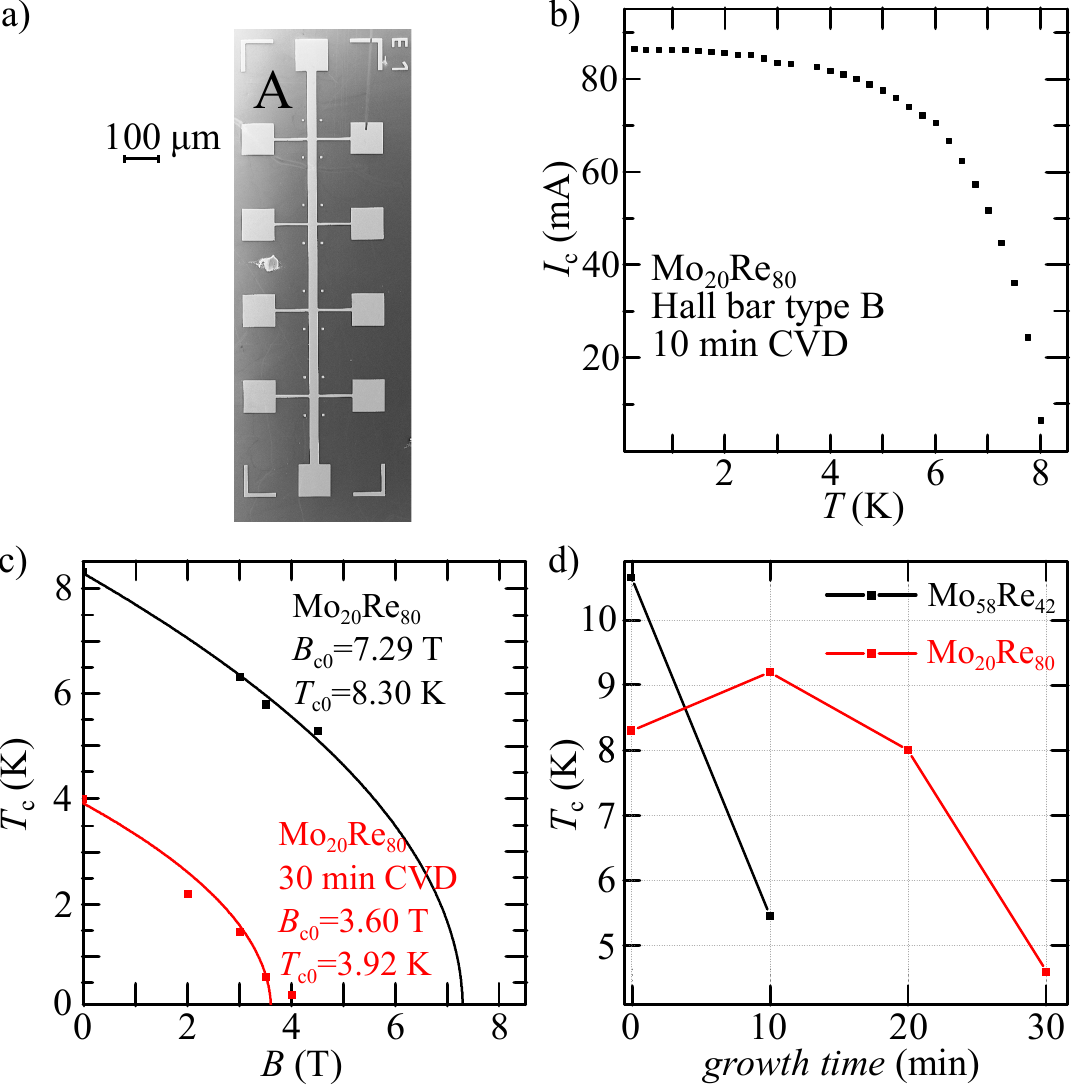} 
\caption{(Color online) (a) Exemplary SEM micrograph of a test Hall bar 
structure, type A. Dimensions are for type A width $W_{\text{A}} = 28 
\mikrom$ and film thickness $d_{\text{A}} \approx 150 \un{nm}$, for type B
$W_{\text{B}} = 5 \mikrom$ and $d_{\text{B}} \approx 60 \un{nm}$. (b) Measured 
critical current $\Ic$ through a Hall bar device (type B), \ReMo{80}{20}, as a 
function of temperature. (c) Critical temperature $\Tc$ of \ReMo{80}{20} films 
as a function of applied magnetic field $B$, with and without CVD growth 
exposure. (d) Measured critical temperature $\Tc$ as a function of $\text{CH}_4 
/ \text{H}_2$ flow time during the CVD process. Straight lines are guides to 
the eye.}\label{fig:4dc}
\end{figure}
To characterize the electronic properties of the co-sputtered films at room 
temperature as well as at cryogenic temperatures, the fabricated thin films on 
top of \SiO or $\text{Al}_{2}\text{O}_{3}$ substrates are patterned into Hall 
bars by means of optical lithography and $\text{SF}_{6} / \text{Ar}$ reactive 
ion etching (see Fig. \ref{fig:4dc}(a)). Afterwards, selected structures are 
placed into the CVD furnace and exposed to the nanotube growth environment for 
several minutes. All measurements have been performed on films either examined 
by XPS (see last two sections) or deposited simultaneously to these in the same 
deposition step. Devices using two different alloy compositions have been 
examined, namely \ReMo{80}{20} and \ReMo{42}{58} obtained with $P_{\text{Mo}} = 
75\un{W}$ and $P_{\text{Mo}}=200\un{W}$, respectively.

At room temperature, compared to \ReMo{42}{58} both resistivity and sheet 
resistance of the \ReMo{80}{20} samples are higher by a factor 3--5: before CVD 
we obtain $\rho\simeq 3.0 \cdot 10^{-7}\,\Omega\text{m}$ for \ReMo{42}{58} and 
$\rho\simeq 9.0 \cdot 10^{-7}\,\Omega\text{m}$ for \ReMo{80}{20}. Resistances 
slightly increase during exposition to the CVD environment, to $\rho\simeq 4.0 
\cdot 10^{-7}\,\Omega\text{m}$ for \ReMo{42}{58} and $13 \cdot 10^{-7} \, 
\Omega\text{m} \lesssim \rho \lesssim 15 \cdot 10^{-7} \, \Omega\text{m}$ for 
\ReMo{80}{20}.

Results of low-temperature measurements performed on \ReMo{80}{20} devices are 
plotted in Fig.~\ref{fig:4dc}(b) and Fig.~\ref{fig:4dc}(c). Independent from 
CVD-exposure, the residual-rest-resistance values $RRR$ for all \ReMo{80}{20} 
devices are in the range $0.8 \lesssim RRR \lesssim 1.0$. Fig.~\ref{fig:4dc}(b) 
displays the critical current of a \ReMo{80}{20} film (Hall bar geometry B, 
with $\Tc = 9.2\un{K}$) after $10 \un{min}$ CVD exposure. It carries a 
supercurrent up to $\Ic \gtrsim 80\un{mA}$, corresponding to a critical current 
density of $\jc \gtrsim 2.7 \cdot 10^5\un{A/mm}^2$ over a wide temperature 
range up to ca. $5\un{K}$ before the transition to a normal conductor takes 
place above $8 \un{K}$. This very high value for $\jc$ is well in accordance 
with the results of \cite{SCsusMoRe}, where after high-temperature 
annealing critical current densities of up to $1.8 \cdot 10^{5} \un{A/mm}^2$ 
through suspended \ReMo{50}{50}-nanostructures were reported.

A second \ReMo{80}{20} Hall bar device (type A, $\Tc=8.3\un{K}$), not exposed 
to CVD, exhibits a critical current $\Ic \approx 114 \un{mA}$ at $T=4.2 
\un{K}$, corresponding to a lower current density $\jc=2.8 \cdot 10^{4} 
\un{A/mm}^2$. Also much longer exposure to the CVD environment again lowers the 
reachable critical current density.

Figure~\ref{fig:4dc}(c) displays data on the magnetic field dependence of the 
critical temperature. Fitting the empirical relation $\Bc(T) = B_{c0}\cdot 
\left( 1 - (T/T_{c0})^{2} \right)$ \cite{Tinkham} results in high 
characteristic values $B_{c0} = 7.3\un{T}$ and $T_{c0}= 8.3\un{K}$ expected for 
molybdenum rhenium alloys, where $T_{c0}$ denotes the zero field critical 
temperature and $B_{c0}$ the extrapolated critical field at zero temperature. 
Again, as observed for critical currents, prolonged ($30 \un{min}$) CVD 
exposure results in a strong decrease of both values to here $B_{c0} = 
3.6\un{T}$ and $T_{c0} = 3.9\un{K}$ \cite{QresonatorsMoRe, SCsusMoRe}. This 
effect is also visible in Figure \ref{fig:4dc}(d), displaying the critical 
temperature for two different alloy compositions. Even for only 10 minutes 
growth time, $\Tc$ of the \ReMo{42}{58} film decreases to the half, well in 
accordance with the results of Ref.~\onlinecite{QresonatorsMoRe}. 
Interestingly, in contrast the \ReMo{80}{20} alloy keeps its critical 
temperature range of $8\un{K}\le \Tc \le 9 \un{K}$ for growth times up to 20 
minutes, only reaching $<4\un{K}$ after 30 minutes of exposure to the 
$\text{CH}_4 / \text{H}_2$ flow.

\section{Coplanar resonator devices}
\begin{figure}[t]
\includegraphics[width=\columnwidth]{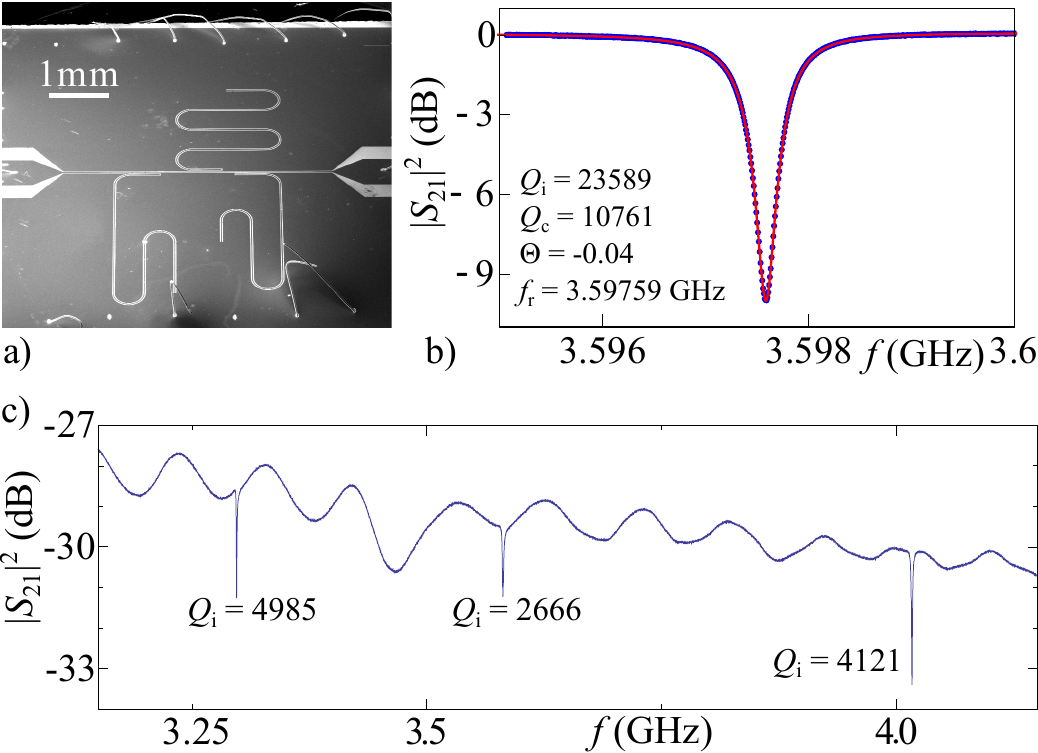} 
\caption{(Color online) (a) SEM image of a test structure with three MoRe 
coplanar $\lambda/4$ resonators coupled to a feed line. (b) Detail of the 
transmission $\left| S_{21}(f) \right|^2$ at $T= 100\un{mK}$ of a device as 
depicted in (a); the background value of $\left|S_{21}(f)\right|^2$ has been 
subtracted. $150\un{nm}$ pristine \ReMo{80}{20}; see the text for a description 
of the fit. (c) Uncalibrated transmission spectrum $\left| S_{21}(f) \right|^2$ 
at $T= 15\un{mK}$ (dilution refrigerator mixing chamber temperature) of a 
second $150\un{nm}$ \ReMo{80}{20} device over a wider frequency range, now 
after $30\un{min}$ exposure to the CVD $\text{CH}_4 / \text{H}_2$ flow. 
Resonances corresponding to the three $\lambda/4$ structures coupled to the 
feedline can still be clearly identified.} 
\label{fig-resonator}
\end{figure}
Coplanar waveguide $\lambda/4$ resonators were fabricated to investigate the 
high frequency behavior of the alloy material and its suitability for cavity 
quantum electrodynamics and optomechanics experiments. After sputtering of 
$150\un{nm}$ \ReMo{80}{20} on \SiO or $\text{Al}_{2}\text{O}_{3}$ substrates, 
the structures were patterned using optical lithography and reactive ion 
etching with $\text{SF}_{6} / \text{Ar}$. A micrograph of a device coupling 
three $\lambda/4$ resonators to a common feed line is shown in 
Fig.~\ref{fig-resonator}(a). The devices are glued on a printed circuit board, 
bonded with aluminum bond wires and subsequently characterized in a dilution 
refrigerator. On the signal input side the experimental wiring of 
superconducting UTF85 semirigid NbTi cables includes attenuators as thermal 
anchoring at every temperature stage. The input signal is attenuated by 
approximately $53\un{dB}$, transmitted through the device under test and then 
amplified by $29\un{dB}$ by a low noise HEMT amplifier\cite{caltech} at the 
$1\un{K}$ stage.

Near its fundamental resonance frequency, given by the length of the resonator 
$l$ and an effective permittivity $\varepsilon_{\text{eff}}$
\begin{equation}
\fr=\frac{1}{\sqrt{\varepsilon_{\text{eff}}}}\frac{c}{4l},
\label{eq:fr}
\end{equation}
each of the three resonators behaves like a parallel lumped-element RLC 
circuit, coupling energy out of the feed line and leading to a distinct 
resonant drop in feedline transmission $S_{21}$. In the vicinity of the 
resonance, the transmitted signal can be expressed by \cite{Khalil2012}
\begin{equation}
S_{21}=1-\frac{\frac{\Ql}{|\Qe|}\cdot \e^{\im \Theta}}{1+2 \im \Ql \cdot 
\frac{f-\fr}{\fr}}, 
\label{eq:S21}
\end{equation}
where $1/\Ql = 1/\Qi + 1/\Qc$. Here, $\Qi$ is the material and temperature 
dependent internal quality factor of the $\lambda/4$ structure, and $\Qc$ the 
geometry dependent coupling quality factor. $\Qe$ is a complex-valued parameter 
closely related to the coupling quality factor $\Qc$, whose finite phase 
$\Theta$ can give rise to a line shape asymmetry due to non-ideal circuit 
elements, e.g. a complex loading of the resonator or impedance mismatches. Its 
real part fulfils the condition $\text{Re}[ \Qe^{-1} ] = \Qc^{-1}$. 
Fig.~\ref{fig-resonator}(b) shows the normalized data of the transmission 
$|S_{21}|^2$ for one exemplary resonance of a pristine film at $T = 100 
\un{mK}$. It includes a fitted curve following Eq.~\ref{eq:S21}; an intrinsic 
quality factor of $\Qi\simeq 23600$ and a  coupling quality factor of $Q_c 
\simeq 10800$ are obtained. Fig.~\ref{fig-resonator}(c) displays an overview 
plot of the uncalibrated transmission of a similar device, this time after an 
exposure of $30\un{min}$ to the hot CVD gas mixture. Still, all three 
resonances can be clearly recognized, with quality factors up to $\Qi\simeq 
5000$. 

\section{Coplanar resonator temperature dependence}
\begin{figure}[t]
\includegraphics[width=\columnwidth]{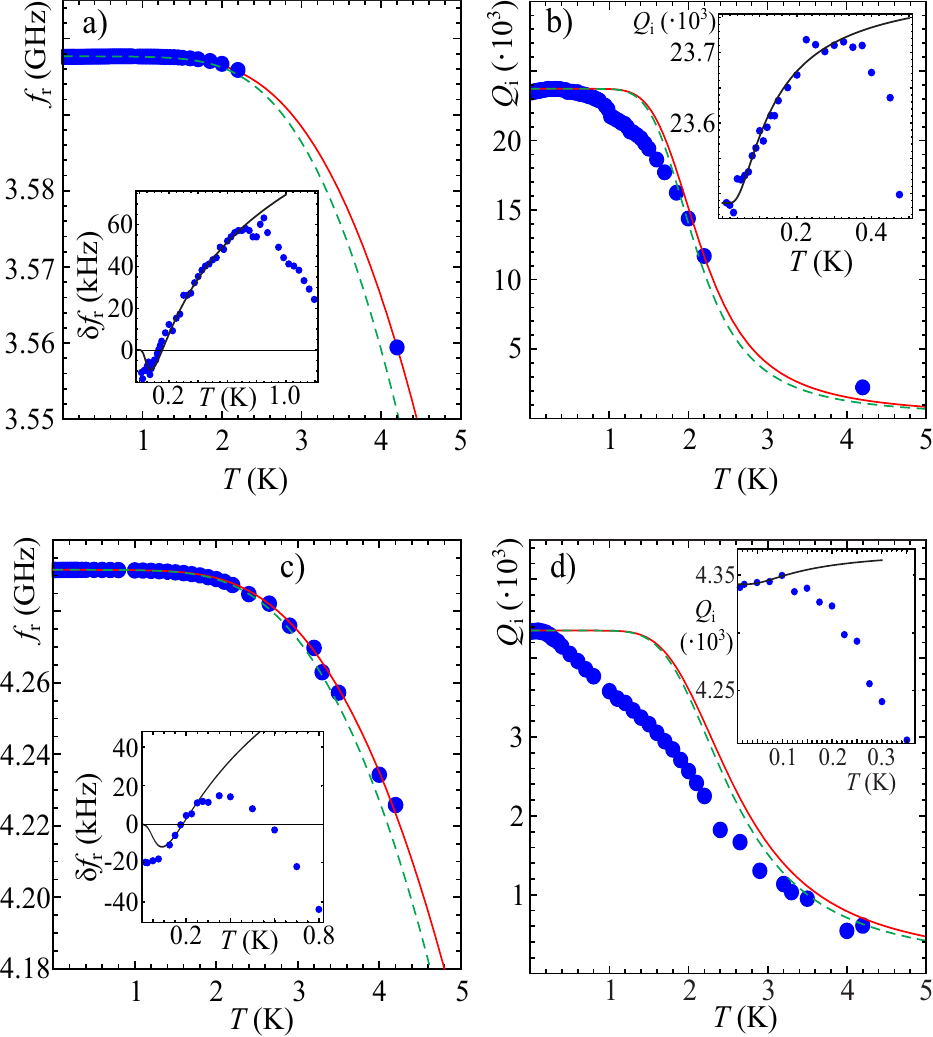} 
\caption{(Color online) Temperature dependence of resonance frequency \fr\ and 
internal quality factor \Qi\ for two $\lambda/4$ resonator devices. (a), (b) 
Resonance frequency \fr\ (a) and internal quality factor \Qi\ (b) for device 1 
(no CVD treatment); compensation-doped silicon substrate covered by 
$500\un{nm}$ thermal $\text{SiO}_2$, $10\un{nm}$ ALD-deposited $\text{Al}_2 
\text{O}_3$ and a $150\un{nm}$ thick \ReMo{80}{20} film. The insets display 
detail zooms for low temperatures. (c), (d) Corresponding plots for device 2, 
after undergoing $10\un{min}$ $\text{CH}_4 / \text{H}_2$ flow in the CVD growth 
oven. Compensation-doped silicon substrate covered by $500\un{nm}$ thermal 
$\text{SiO}_2$ and $150\un{nm}$ \ReMo{80}{20}. For a description of the fit 
models (solid and dashed lines) see the text.}
\label{fig-resonator-temp}
\end{figure}
Fig.~\ref{fig-resonator-temp} shows the observed temperature dependence of the 
resonance frequency $\fr(T)$ (Figs.~\ref{fig-resonator-temp}(a,c)) and the 
internal quality factor $\Qi(T)$ (Figs.~\ref{fig-resonator-temp}(b,d)) for two 
devices. The device of Figs.~\ref{fig-resonator-temp}(a,b) (device 1) has been 
characterized after thin film deposition and patterning, the device of 
Fig.~\ref{fig-resonator-temp}(c,d) (device 2) has been additionally exposed to 
the nanotube CVD growth environment. 

As can be seen in the figure, the resonance frequency $\fr$ clearly decreases 
at high temperatures ($T\gtrsim 0.8\un{K}$ for device 1, $T\gtrsim 0.4\un{K}$ 
for device 2). This can be attributed to a decrease in superfluid density of the 
superconducting thin film. A related rise in the quasiparticle density leads to 
a higher damping of the resonator and therefore to a reduction of the internal 
quality factor. Following Mattis and Bardeen \cite{MB1958, QresonatorsMoRe}, 
the temperature dependence of \fr\ can be expressed as
\begin{equation}\label{eq:freq-mb}
\frac{\delta \fr}{f_0}  = \frac{\alpha_0}{2} \, \frac{\delta 
\sigma_{2}}{\sigma_{2}}, 
\end{equation}
with $\delta \fr(T) = \fr(T) - f_{0}$ the deviation of the resonance frequency 
with finite temperature and $f_{0}=\fr(T=0)$ the approximated resonance 
frequency at $T=0$. In the limit $h f \ll \Delta(T=0)$ and $\kB T \ll 
\Delta(T=0)$ the imaginary part $\sigma_{2}$ of the complex conductivity
$\sigma$ of the device can be approximated using the temperature-dependent BCS 
energy gap $\Delta$ and the normal state conductivity $\sigma_n$ 
as \cite{GaoZmuidzinas2008}
\begin{equation}
 \frac{\sigma_2}{\sigma_n}=
 \frac{\uppi \Delta}{h f}
 \left[
 1-2\e^{-\Delta/\kB T} \e^{-h f/2 \kB T} I_0\left(\frac{h f}{2 \kB T}\right)
 \right]
 ,
\end{equation}
where $I_0(x)$ is a modified Bessel function of the first kind. The parameter 
$\alpha_0$ in Eq.~\ref{eq:freq-mb} is the kinetic inductance fraction in 
zero-temperature limit of the coplanar waveguide. 

The solid red lines in Figs.~\ref{fig-resonator-temp}(a,c) are fit curves
corresponding to this model, using in each case $\alpha_0$ as a free parameter. 
The result agrees very well with the experimental data. For device 1 (no CVD) we 
obtain $\alpha_0 = 0.199$, for device 2 (after $10\un{min}$ CVD) $\alpha_0 = 
0.249$. The value of $\alpha_0$ can be calculated from the normal-state 
conductance $\sigma_n$ and the critical temperature \Tc\ following 
\cite{collin}. Using the parameters of our devices, we obtain 
$\alpha_\text{th}=0.243$ and $\alpha_\text{th}=0.284$, in reasonable agreement 
with the fit results. The corresponding functional dependence of $\fr(T)$ is 
plotted in Figs.~\ref{fig-resonator-temp}(a,c) each as a green dashed line. 

In Figs.~\ref{fig-resonator-temp}(b,d), clearly also the internal quality 
factor $Q_i$ of the devices decreases at high temperature. Following Mattis and 
Bardeen, the corresponding change of the quality factor is
\begin{equation}
 \delta\left( \frac{1}{\Qi} \right) = \alpha_0 \frac{\delta \sigma_1}{\sigma_2}
\end{equation}
with the real part of the complex conductivity \cite{GaoZmuidzinas2008}
\begin{equation}
 \frac{\sigma_1}{\sigma_n}= \frac{4\Delta}{h f} 
 \e^{-\Delta/\kB T} 
 \text{sinh}\left(\frac{h f}{2 \kB T}\right)
 K_0 \left( \frac{h f}{2 \kB T}\right).
\end{equation}
Here $K_0(x)$ is a modified Bessel function of the second kind. Using the 
parameters extracted in Figs.~\ref{fig-resonator-temp}(a,c) we can plot the 
corresponding expected temperature dependence $\Qi(T)$, again as solid red and 
dashed green lines; while significant deviations exist, the overall tendency 
agrees well with the data. 

For $T \ll \Tc$, the theory by Mattis and Bardeen predicts no change in the 
resonance frequency with temperature. However, in our devices for $T\lesssim 
0.5\un{K}$  a slight decrease in $\fr$ is observed, leading to an overall 
nonmonotonic behaviour of $f_r(T)$, see the insets of 
Figs.~\ref{fig-resonator-temp}(a,c). This is consistent with the influence of 
two-level systems (TLS) in the substrate contributing to both dissipation and 
dispersion and can be described by \cite{Phillips1987, Bruno2014, 
Pappas2011} 
\begin{align}
\frac{\delta \fr}{f_{0}} &= \frac{F}{2} \frac{\delta \varepsilon}{\varepsilon} 
\\ &= 
\frac{F\, \vartheta}{\uppi}\left[\text{Re}\Psi 
\left(\frac{1}{2}+\frac{1}{2\uppi\,\im}\frac{h 
\fr(T)}{k_{\text{B}}T}\right) - \text{ln}\left(\frac{1}{2\uppi}\frac{h 
\fr(T)}{k_{\text{B}}T}\right) \right].
\end{align}
Here, $\Psi$ is the digamma function, $0<F<1$ the filling factor, giving the 
ratio of the electric energy stored in the TLS hosting material to the total 
electric energy stored in the resonator, and $\vartheta$ the loss tangent of 
the substrate. The insets of Figs.~\ref{fig-resonator-temp}(a,c) show the 
measured data along with a low-temperature fit, using the product 
$F\,\vartheta$ and $f_{0}$ as fit parameters. 

In the case of device 1, Fig.~\ref{fig-resonator-temp}(a), this results in 
$F\,\vartheta = 3.968\cdot 10^{-5}$. This value of $F\,\vartheta$ is comparable 
to literature values for niobium resonators on sapphire, see e.g. 
\cite{Gao2008}. The fit value for $F\,\vartheta$ provides an approximation for 
the TLS-related low-power, low-temperature internal Q-factor $Q_{\text{i,TLS}} 
\approx 1/(F\, \vartheta) \approx 25000$. As expected this slightly exceeds our 
measured value of $\Qi(T=20\un{mK}) \approx 23500$. The fit for device 2 
provides $F\,\vartheta = 4.087\cdot 10^{-5}$. This results in a quality factor 
$Q_{\text{i,TLS}} \approx 24500$, comparable to the value for device 1 and 
apparently insensitive to the degradation of the device due to the CVD process.

Similar to $f_0$, also the low-temperature behavior of the quality factor $\Qi$ 
is governed by interactions with substrate TLS. This can be modeled by 
\cite{Bruno2014, Pappas2011}
\begin{equation}
\frac{1}{\Qi}=F \, \vartheta_{\text{eff}}\, \tanh \left(
\frac{h\fr(T)}{2k_{\text{B}}T}
\right)+\frac{1}{Q_{\text{other}}}.
\end{equation}
Here, $ \vartheta_{\text{eff}}$ is an effective, reduced loss tangent which 
takes into account that at strong driving the two-level systems are partially 
saturated and thereby unable to absorb energy. $Q_{\text{other}}$ describes
dissipative processes unrelated to TLS. The fit for device 1 again agrees well 
with the measured data (inset of Fig. \ref{fig-resonator-temp}(b)) and results 
in $F\,\vartheta_{\text{eff}} = 5.677\cdot 10^{-7}$ and $Q_{\text{other}} = 
23800$. For device 2 (inset of Fig. \ref{fig-resonator-temp}(d)) only few data 
points at low temperature are available, and a saturation of the device 
temperature cannot be excluded. Applying the fit, we obtain $F \, 
\vartheta_{\text{eff}} = 1.634\cdot 10^{-6}$ and $Q_{\text{other}} = 4370$, 
dominating the TLS contribution.

As can be seen in Fig.~\ref{fig-resonator-temp}, the combined effect of 
two-level systems and a temperature dependent superfluid density provide a good 
description of our devices. During the CVD process the kinetic inductance 
fraction strongly increases. The low temperature fits tentatively display 
similar influence of two-level systems before and after CVD, however a 
significant non-TLS induced dissipation term results in the second, post-CVD 
device.

\section{Summary and conclusions}

The x-ray photoelectron spectroscopy characterization demonstrates that our
co-sputtering process from two independent targets can generate MoRe thin films 
of controlled alloy composition. We observe traces of molybdenum oxides 
at the film surface after ambient air exposure. Exposing the thin films to the 
carbon nanotube CVD growth environment, a significant amount of carbon is 
incorporated into the film. No clear indications for the formation of 
molybdenum carbide can be found; the surface oxide is absent after CVD, 
pointing towards its reduction in the $850\,^\circ\text{C}$ $\text{H}_2$ 
atmosphere.

In electrical characterization, we observe a higher resilience to the CVD 
process for $\ReMo{80}{20}$ films; the critical temperature only drops below 
$\Tc=8\un{K}$ for growth times $>20\min$. For shorter growth times, data 
indicates that the CVD process may have effects similar to the annealing 
discussed in \cite{SCsusMoRe}, i.e., enhancing the critical temperature and 
critical current density. We observe up to $\jc \simeq 2.7\cdot 
10^5\un{A/mm}^2$ and $\Tc \simeq 9.2\un{K}$ in a Hall bar geometry. 

$\lambda/4$ coplanar waveguide resonators were defined by thin film deposition 
and subsequent reactive ion etching. After structuring, we observe internal 
quality factors up to $\Qi=23700$ in pristine devices and up to $\Qi=5000$ 
after undergoing the CVD process. The temperature evolution of the resonance 
frequency \fr\ and the internal quality factor \Qi\ of the devices can be 
understood for low temperatures $T\lesssim 0.4\un{K}$ via interaction with 
substrate two-level systems, at higher temperatures $T\gtrsim 0.8\un{K}$ 
via the decreasing superfluid density.

\ReMo{80}{20}\ clearly excels in terms of stability during the CVD process, 
critical temperature and field as well as critical current. However, the
achieved quality factors are clearly lower than those reported in literature 
for similar materials \cite{QresonatorsMoRe}. Our reference niobium devices 
characterized for comparison have reached $\Qi\simeq 4\cdot 10^5$, excluding 
the detection setup as cause. The effect of the CVD process in lowering $\Qi$ 
is similar to other published observations \cite{QresonatorsMoRe}. Further 
improvements thus should be targeted at the radiofrequency properties of the 
substrate as well as the the pristine, as-deposited metal film and coplanar 
waveguide resonators defined in it.

\section{Acknowledgments}
The authors gratefully acknowledge funding by the Deutsche 
Forschungsgemeinschaft via SFB 631, GRK 1570, and Emmy Noether project 
Hu~1808/1.

\bibliography{paper}

\end{document}